%% file: ELRwtE.tex
\documentclass{article}

\usepackage{url}
\usepackage{listings}

\usepackage{common}

\begin{document}

\title{\normalfont\bfseries\large Eliminating Left Recursion without the Epsilon}
\author{James Smith\\\texttt{james.smith@djalbat.com}}
\date{}
	
\maketitle

\begin{abstract}
\noindent The standard algorithm to eliminate indirect left recursion takes a preventative approach, rewriting a grammar's rules so that indirect left recursion is no longer possible, rather than eliminating it only as and when it occurs. This approach results in many of the rules being lost, so that the parse trees that result are often devoid of the detail that the BNF was supposed to capture in the first place. Furthermore, the standard algorithm results in exponential blow-up as the BNF is rewritten, making it wholly unworkable in practice. To avoid these pitfalls, we revise the standard algorithm to eliminate direct left recursion and then take a graph-theoretic approach to eliminating indirect left recursion. We also extend the algorithm to rewrite the resultant parse trees in order to recover the parse trees that would have resulted if left recursion had not had to be eliminated in the first place. Therefore, aside from a couple of caveats, our algorithm works not just in theory but also in practice.
\end{abstract}

\include*{introduction}
\include*{direct}
\include*{indirect}
\include*{rewriting}

\include*{conclusions}

\pagebreak

\bibliographystyle{plain}
\bibliography{references}

\end{document}

%% file: introduction.tex
\section{Introduction}

Top-down parsers seemingly have several advantages. They are easy to understand, obvious even, and thus relatively easy to implement. They are also comparatively fast and amenable to incremental parsing. However, they do have an Achilles heel, namely left recursion, which causes them to loop indefinitely whenever they encounter it. Consider listing~\ref{rules-for-simple-expressions}, which shows a collection of rules in Backus–Naur form~\cite{Wikipedia:BNF} suitable for parsing simple arithmetic expressions such as $(1+2)/3$.

\begin{lstlisting}[caption={Rules for parsing simple arithmetic expressions},label={rules-for-simple-expressions}]
expression    ::= expression operator expression

                | "(" expression ")"
             
                | term
             
                ;

operator      ::= "+" | "-" | "/" | "*" ;

term          ::= naturalNumber ;

naturalNumber ::= /\d+/ ;
\end{lstlisting}
When a top-down parser attempts to execute a rule, it tries each definition in turn until it finds one that can be fully executed, that is, each of its parts can be executed in sequence. In this case, a top-down parser would try the first definition of the starting {\small\texttt{expression}} rule and would immediately encounter a reference to that rule, which it would dutifully try to execute again and hence begin to loop indefinitely. In such cases we say that the definition is directly left recursive. Left recursion results only when the first part of a definition references a rule, by the way, not any other part. The second definition is not left recursive, for example, because a top-down parser would try the terminal symbol part first, and would therefore have the opportunity of either executing that part or of trying another definition if that part did not execute.

A directly left recursive definition can easily be dealt with by removing it, generating a corresponding right recursive rule and then adding references to this rule to all of the original rule's remaining definitions. Consider listing~\ref{expression-rule-with-left-recursion-eliminated},\footnote{All of the listings in this paper can be found in the readme file of the accompanying implementation~\cite{Occam:Grammar-Utilities}. It also has an example application that can be run directly from a browser, which allows each of the listings to be tried directly. It is much easier to copy them from the implementation's readme file than from this paper, by the way. We think that the parse trees that the example application shows have a certain aesthetic that is worth a little effort.} where the previous {\small\texttt{expression}} rule has been rewritten in this way. Note that the right recursive {\small\texttt{expression\textasciitilde}} rule has a second definition with a single $\epsilon$ part, which we call a vacuous definition. When this part is encountered, it is always trivially executed, meaning that the definition and therefore the rule as a whole executes even though no tokens are consumed.

\begin{lstlisting}[caption={The expression rule with left recursion eliminated},label={expression-rule-with-left-recursion-eliminated}]
expression  ::= "(" expression ")" expression~
             
              | term expression~
             
              ;

expression~ ::= operator expression expression~

              | \epsilon
             
              ;
\end{lstlisting}
Now consider listing~\ref{rules-for-simple-expressions-with-indirect-left-recursion}, which is the result of rewriting the first rule of listing~\ref{rules-for-simple-expressions}, removing its first definition and placing it in its own rule. This results in what we call indirect left recursion. We can see that if the parser tries to execute the first {\small\texttt{expression}} rule it will come across a reference to the second {\small\texttt{compoundExpression}} rule. It will duly try to execute that rule, which entails trying to execute its sole definition. However, the first part of that definition references the {\small\texttt{expression}} rule again, and thus the parser will once again loop indefinitely. In this case we say that the first definition of the {\small\texttt{compoundExpression}} rule is indirectly left recursive. The first definition of the first {\small\texttt{expression}} rule, on the other hand, which bares some of the blame for the left recursion so to speak, we call implicitly left recursive.\footnote{We have chosen slightly non-standard terms for some of the types of left recursion. What is generally referred to as immediate left recursion, for example, we call direct left recursion. And what is generally called implicit left recursion, we call indirect left recursion. We will define all of these definitions formally later on.}

\begin{lstlisting}[caption={Rules for parsing simple arithmetic expressions with indirect left recursion},label={rules-for-simple-expressions-with-indirect-left-recursion}]
expression         ::= compoundExpression

                     | "(" expression ")"
             
                     | term
             
                     ;

compoundExpression ::= expression operator expression ;
\end{lstlisting}
In contrast to direct left recursion, indirect left recursion has traditionally been eliminated by an algorithm that works preventatively~\cite{Power:Eliminating-Indirect-Left-Recursion}. It does not search for indirect left recursion and eliminate it explicitly, instead it reorganises the rules in such a way that indirect left recursion is no longer possible. To give an idea, the definitions of the {\small\texttt{compoundExpression}} rule would be substituted for the reference to it in the first definition of the {\small\texttt{expression}} rule, effectively recovering the {\small\texttt{expression}} rule defined initially in listing~\ref{rules-for-simple-expressions}. By making these repeated substitutions, the algorithm guarantees that if there is any indirect left recursion then it will eventually surface as direct left recursion, in which case it is eliminated by the standard algorithm. Therefore direct left recursion, which can be thought as a degenerate case of indirect left recursion, is guaranteed to be eliminated, too. 

The problem with this approach, however, is that the {\small\texttt{compoundExpression}} rule, or the fact that it is, or rather was, referenced by the {\small\texttt{expression}} rule, is lost for good. It is these relationships between roles that give parse trees their utility in practice, however, and we cannot afford to simply throw them away. Furthermore, it is easy to see that if such substitutions are done on many occasions, we may get exponential blow up both in the number of definitions overall and in their length. Rules that create such blow ups are alarmingly commonplace in practice, in fact, and result in the parser simply being unable to function without running out of stack or heap space.

So can we rewrite the indirectly left recursive definition in such as way as to leave its relation to the {\small\texttt{expression}} rule intact? If the answer to that question is yes, it seems not implausible that the process could be extended to eliminate indirect left recursion in most if not all cases. Fortunately, it turns out that the answer is yes. What helped in finding the answer was that the algorithm to eliminate direct left recursion was revised first, in order to accommodate some features of the variant of extended BNF that we employ, and this revision led in turn to some unexpected gains when tackling the more difficult problem. Therefore we begin by explaining how the algorithm to eliminate direct left recursion was revised, and then we go on to show how it was extended to an algorithm to eliminate the indirect form, and therefore all forms, of left recursion.

%% file: direct.tex
\section{Eliminating direct left recursion} 

\noindent Listing~\ref{direct-left-recursion-simplest-case} shows the simplest case of direct left recursion.  

\begin{lstlisting}[caption={Simplest case of direct left recursion},label={direct-left-recursion-simplest-case}]
A ::= A "f" 

    | "g"   

    ;   
\end{lstlisting}

\noindent It is reasonable to ask, what does left recursion buy us? The answer is repetition. Is should be clear that the sentences that these rules match, assuming of course that our parser was not susceptible to left recursion,  would be the letter `g' followed by zero or more letter `f's. It should also be clear that the rules in listing~\ref{direct-left-recursion-simplest-case} can be written as in listing~\ref{rewritten-direct-left-recursion-simplest-case}. Thus we take a different approach to the standard algorithm, which rewrites left recursion as right recursion.  Our variant of BNF supports a `*' zero or more qualifier, so we employ that.

\begin{lstlisting}[caption={Rewritten simplest case of direct left recursion},label={rewritten-direct-left-recursion-simplest-case}]
A ::= "g" "f"* ;   
\end{lstlisting}

\noindent Listing~\ref{direct-left-recursion-more-complex-case} shows a more complex case of direct left recursion.

\begin{lstlisting}[caption={More complex case of direct left recursion},label={direct-left-recursion-more-complex-case}]
A ::= A "f" "g"  

    | A "h" 

    | "k"     

    | "j"      

    ;           
\end{lstlisting}

\noindent Its elimination follows the same principles, however, with the result being shown in listing~\ref{rewritten-direct-left-recursion-more-complex-case}. Note that in this case two additional {\small\texttt{A\_}} and {\small\texttt{A\textasciitilde}} rules have been created, which we call the reduced and directly repeated rules, respectively, for hopefully obvious reasons. In fact all cases of direct left recursion can be rewritten in this way,, unless they are degenerate cases, by which we mean that even a parser that is not susceptible to left recursion would be unable to execute them; or they cannot be rewritten for a few other good reasons.  We cover these as special cases next.

\pagebreak

\begin{lstlisting}[caption={Rewritten more complex case of direct left recursion},label={rewritten-direct-left-recursion-more-complex-case}]
A  ::= A_ A~* ;

A_ ::= "k"
 
     | "j"         
    
     ;

A~ ::= "f" "g"
    
     | "h"
                  
     ;
\end{lstlisting}

\subsection{Special cases} 

The first special case, an example of which is shown in listing~\ref{direct-left-recursion-complex-part-special-case}, concerns what are called complex parts. Like the repetition quantifier already covered, these are features of the variant of BNF that we are using.

\begin{lstlisting}[caption={A special case of complex part inhibiting the rewriting of direct left recursion},label={direct-left-recursion-complex-part-special-case}]
A ::= "f"

    | ( A | B ) "g"

    | "h"

    ;
\end{lstlisting}

\noindent Here the first part of the second definition, called a choice of parts part, is directly left recursive. Such parts are called complex because they cannot be easily rewritten. It is possible to rewrite them, but doing so would mean creating an intermediate rule, and would therefore add considerably to the complexity of the algorithm. So for its part the algorithm throws an error suggesting that the user pull out the offending part into a separate rule. This gives the user a chance to choose a suitable name for that rule to boot, rather than the algorithm providing a manufactured one.

Moving on, an example of the next special case is shown in listing~\ref{empty-repeated-rule-special-case}. Here the repeated rule would be empty and therefore even if the parser was not susceptible to left recursion, it would still not be able to execute the definition. Note that in such cases a fault with the BNF can be clearly identified by a property of a repeated rule.

\begin{lstlisting}[caption={A special case of an empty repeated rule},label={empty-repeated-rule-special-case}]
A ::= "g"

    | A

    | "f"

    ;
\end{lstlisting}

\noindent In fact there is a variant of this special case, due to what we call non-consuming definitions. We say that a definition is non-consuming if it can be executed any tokens being consumed. A directly repeated rule cannot be non-consuming because if that were the case then it could be executed indefinitely. Listing~\ref{effectively-empty-directly-repeated-definition-special-case} gives an example.

\pagebreak

\begin{lstlisting}[caption={A special case of an non-consuming directly repeated definition},label={effectively-empty-directly-repeated-definition-special-case}]
A  ::= A_ A~* ;
      
A_ ::= "f" ;
    
A~ ::= "g"? ;
\end{lstlisting}

\noindent In a similar vein, when the initial parts of a definition are non-consuming the first part that is not non-consuming effectively becomes left recursive. These cases, which we call occluded left recursion, are ruled out with an exception. In passing it is also worth noting that whether or not a part is non-consuming is decided recursively. Imagine a rule name part that references a rule that is non-consuming,  for example.

As a problem with the BNF can be identified as a property of a directly repeated rule, namely that it is non-consuming, so an empty reduced rule is also an indicator of a problem. This happens when all the definitions of a rule are directly left recursive, an example of which is shown in listing~\ref{two-isolated-directly-left-recursive-definitions-special-case}. In these cases the algorithm again throws an error as there is nothing to be done but to rethink the original rule. In short, all directly left recursive rules need to include at least one non-directly left recursive definition regardless of whether the parser is susceptible to left recursion or not.

\begin{lstlisting}[caption={A special case of two isolated directly left recursive definitions},label={two-isolated-directly-left-recursive-definitions-special-case}]
A ::= A B 

    | A "c"
    
    ;
\end{lstlisting}

\noindent The last of the special cases concerns quantifiers and modifiers applied to directly left recursive parts, an example of which is shown in listing~\ref{directly-left-recursive-definition-with-quantifier}.

\begin{lstlisting}[caption={A special case of a directly left recursive definition with a quantified first part},label={directly-left-recursive-definition-with-quantifier}]
A ::= A? "f" "g"
    
    | "d"
    
    ;
\end{lstlisting}

\noindent The reader might ask, how does the zero or more quantifier on the first part relate to the repetition resulting from the direct left recursion? It turns out that this is not such an easy question to answer. In fact, there appears to be no answer to it at all, at least in the context of the elimination of indirect left recursion, where the rewriting that has been used so far has to be subtly augmented. Therefore the algorithm throws an error and we leave it at that.

%% file: indirect.tex
\section{Eliminating indirect left recursion}

\noindent Listing~\ref{indirect-left-recursion-another-simple-case} shows another simple case of indirect left recursion.  

\begin{lstlisting}[caption={Another simple case of indirect left recursion},label={indirect-left-recursion-another-simple-case}]
A ::= B "f" 

    | "g"   

    ;   

B ::= B "k" 

    | A "h"   

    ;   
\end{lstlisting}

\begin{figure}[H]
\centering
\includegraphics[scale=0.5]{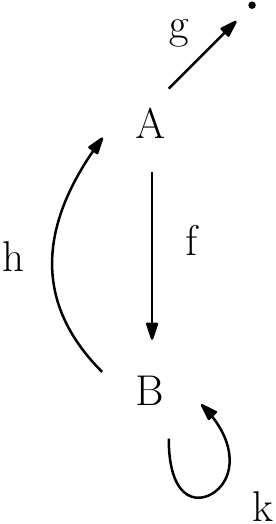} 
\caption{Another simple case of indirect left recursion}
\label{indirect-left-recursion-another-simple-case-graph}
\end{figure}

\noindent We represent this BNF in the form of a directed graph as shown in figure~\ref{indirect-left-recursion-another-simple-case-graph}. With a little practice it is easy to traverse this graph in order to compose the sentences that the corresponding BNF matches. The only slightly counter-intuitive part is that we have to compose these sentences from right to left. For example, the aforementioned BNF will match the following sentence:

\begin{center}
\begin{tabular}{c}
ghfhkkf
\end{tabular}
\end{center}

\noindent To justify this we offer the following derivation:

\begin{center}
A \textasciitilde ghfhkkf  -> B \textasciitilde ghfhkk  -> B \textasciitilde ghfhk  -> B \textasciitilde ghfh -> A \textasciitilde ghf  -> \textasciitilde gh  ->  \textasciitilde g 
\end{center}

\noindent Here we have both direct and indirect left recursion. The reasoning above suggests that they can both be handled and listing~\ref{indirectly-repeated-rules-another-simple-case} shows what we call the indirectly repeated rules that will be needed. Note the presence of the {\small\texttt{A\textasciitilde A}} rule.  Despite its one definition being vacuous, as we say, it cannot be omitted, for reasons that will become apparent. Note also that these rules allow for multiplicity in the indirectly left recursive definitions analogously to direct left recursion.

\begin{lstlisting}[caption={Indirectly repeated rules for another simple case of indirect left recursion},label={indirectly-repeated-rules-another-simple-case}]
A~A ::= \epsilon ;

A~B ::= "f" ;   

B~B ::= "k" ;   

B~A ::= "h" ;   
\end{lstlisting}

\noindent In fact the reduced rules are created and the existing rules rewritten in much the same way as before. They are shown in listing~\ref{reduced-and-rewritten-rules-another-simple-case}.  There are two points worth noting. Firstly,  the reduced {\small\texttt{B\_}} rule has been omitted because it would be empty and therefore the original {\small\texttt{B}} rule has also been rewritten to a shortened form. Some may wonder why this does not create a problem. The answer is that only one rule in any particular indirectly left recursive cycle needs to allow the parser a chance to terminate. Since the {\small\texttt{A}} rule in this case has a non-left recursive definition, we are fine. Secondly, the {\small\texttt{B}} rule contains a single definition that admits a path back from the {\small\texttt{B}} to the {\small\texttt{A}} node. Only by executing this rule can the parser ever terminate and therefore such a path has to added by way of this definition.

\begin{lstlisting}[caption={Reduced and rewritten rules for another simple case of indirect left recursion},label={reduced-and-rewritten-rules-another-simple-case}]
A  ::= A_ A~* ;

B  ::= A_ A~* B~A ;   

A_ ::= "g" ;   
\end{lstlisting}

\begin{figure}[H]
\centering
\includegraphics[scale=0.5]{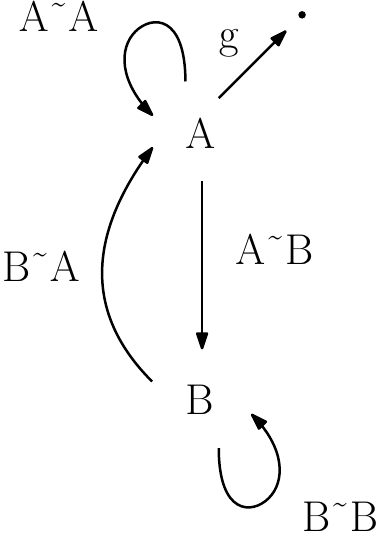} 
\caption{Another simple case of indirect left recursion rewritten}
\label{rewritten-indirect-left-recursion-another-simple-case-graph}
\end{figure}

\noindent This second point gives a clue as to how to define the {\small\texttt{A\textasciitilde}} and {\small\texttt{B\textasciitilde}} directly repeated rules. In order to see how this can be done precisely, we rewrite the previous graph to highlight the indirectly repeated rules. This rewritten graph is shown in figure~\ref{rewritten-indirect-left-recursion-another-simple-case-graph}. Each path must correspond to a definition in one of these rules. Starting with the {\small\texttt{A\textasciitilde}} directly repeated rule, it must include as a definition any path that starts at the {\small\texttt{A}} node and ends there. There are two such paths, namely the single path with the {\small\texttt{A\textasciitilde A}} label and the composite path that passes through the {\small\texttt{B}} node.  Similarly for the {\small\texttt{B\textasciitilde}} rule, which must contain as a definition any path that starts at the {\small\texttt{B}} node and ends there.  Perhaps it is best just to show the definitions in listing~\ref{directly-repeated-rules-another-simple-case}.

\begin{lstlisting}[caption={Directly repeated rules for another simple case of indirect left recursion},label={directly-repeated-rules-another-simple-case}]
A~ ::= A~A

     | B~A B~* A~B 
    
     ;

B~ ::= B~B 

     | A~B A~* B~A
       
     ;   
\end{lstlisting}

\noindent There is a subtlety in these rules, namely that as the two composite paths pass through their opposing nodes they must be allowed to linger, so to speak, tracing out as many circular paths on that node as necessary. Hence the definitions are mutually recursive but never, thankfully, mutually left recursive. One other subtlety deserves mention, and that is the ordering of the parts in the definitions. Looking at the graph, they read from right to left as the paths are traversed. It is worth a moment to gain an assurance that this is indeed the correct ordering.

All of the rules can be seen in listing~\ref{all-rules-another-simple-case} and there is final point worth making, namely the inclusion of a start rule that has been omitted up until now. This start rule has two parts. The first part is a rule name part with a look-ahead modifier. This modifier ensures that the parser will try all the available combinations of rules, definitions and parts until it finds one that allows it to proceed to the second part. This second part is a terminal part that matches an end of line character, but it could be any part. 

By switching the parser to look-ahead mode in his way, we can guarantee that our newly created and rewritten rules will indeed find a path through the graph that consumes the given sentence, if such a path exists. Without the look-ahead modifier, it is highly unlike that this path would be found. It is not a good idea to leave the parser in look-ahead mode continually because it will perform so much more slowly and for this reason look-ahead modifiers should be use sparingly, if at all. However, they are unavoidable in these cases, unfortunately, and we take the performance hit in return for the huge usability gain that eliminating left recursion brings.

\pagebreak

\begin{lstlisting}[caption={All the rules for another simple case of indirect left recursion},label={all-rules-another-simple-case}]
S   ::= A... <END_OF_LINE> ;

A   ::= A_ A~* ;

B   ::= A_ A~* B~A ;   

A_  ::= "g" ;   

A~  ::= A~A

      | B~A B~* A~B 
    
      ;

B~  ::= B~B 

      | A~B A~* B~A
       
      ;   

A~A ::= \epsilon ;

A~B ::= "f" ;   

B~B ::= "k" ;   

B~A ::= "h" ;   
\end{lstlisting}

\begin{figure}[H]
\centering
\includegraphics[scale=0.75]{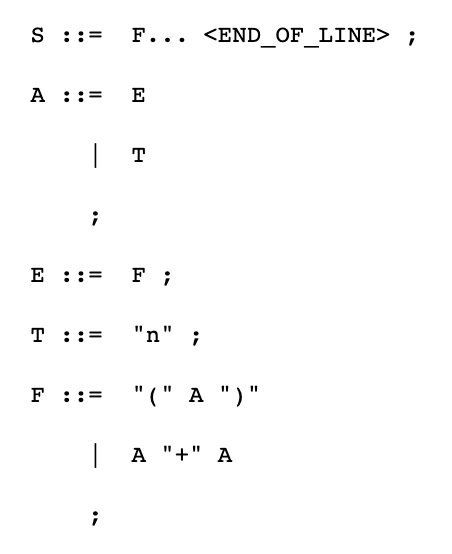} 
\caption{Complex indirect left recursion rules}
\label{complex-indirect-left-recursion-rules}
\end{figure}

\noindent We end this section with a more complex example of indirect left recursion. The rules are shown in figure~\ref{complex-indirect-left-recursion-rules}. They come from a real world case with, for example, the {\small\texttt{A}},  {\small\texttt{T}}  and  {\small\texttt{E}}  rules corresponding to whittled down rules for arguments, terms and expressions.

\begin{figure}[H]
\centering
\includegraphics[scale=0.5]{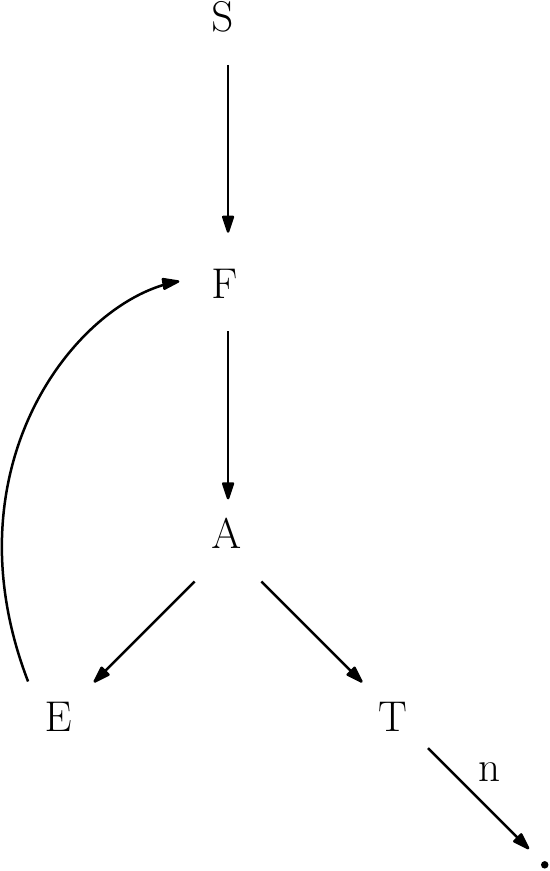} 
\caption{Complex indirect left recursion graph}
\label{complex-indirect-left-recursion-graph}
\end{figure}

\noindent These rules were one of many special cases that had to be handled, and caused considerable difficultly, before the graph-theoretic approach was adopted. The corresponding graph is shown in figure~\ref{complex-indirect-left-recursion-graph}. Note that there is only one non-trivial left recursive cycle. 

\begin{figure}[H]
\centering
\includegraphics[scale=0.75]{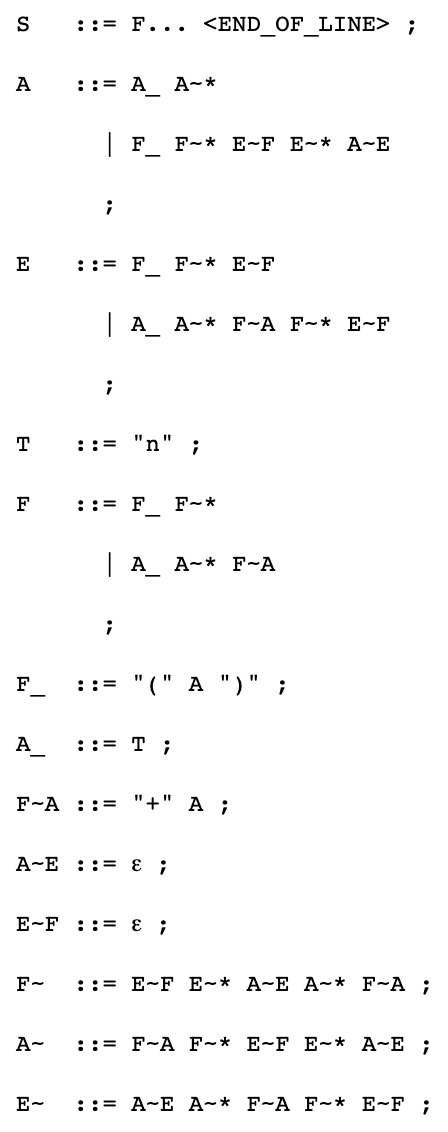} 
\caption{Complex indirect left recursion rewritten rules}
\label{complex-indirect-left-recursion-rewritten-rules}
\end{figure}

The reduced, repeated and rewritten rules after the algorithm has been run are shown in figure~\ref{complex-indirect-left-recursion-rewritten-rules}. This more complex case and the inclusion of this listing in particular are not meant to bamboozle, by the way. On the contrary, the listing is meant to show how relatively easy it is to deal with more complex cases. Note the definitions in the directly repeated rules in particular. Again they describe the one non-trivial left recursive cycle and include the customary mutual recursion. The rewritten  {\small\texttt{A}},  {\small\texttt{E}}  and  {\small\texttt{F}} rules also deserve attention, with their definitions describing the requisite paths from one node to another in the graph that ensure that the parser always terminates when necessary.

\begin{figure}[H]
\centering
\includegraphics[scale=0.5]{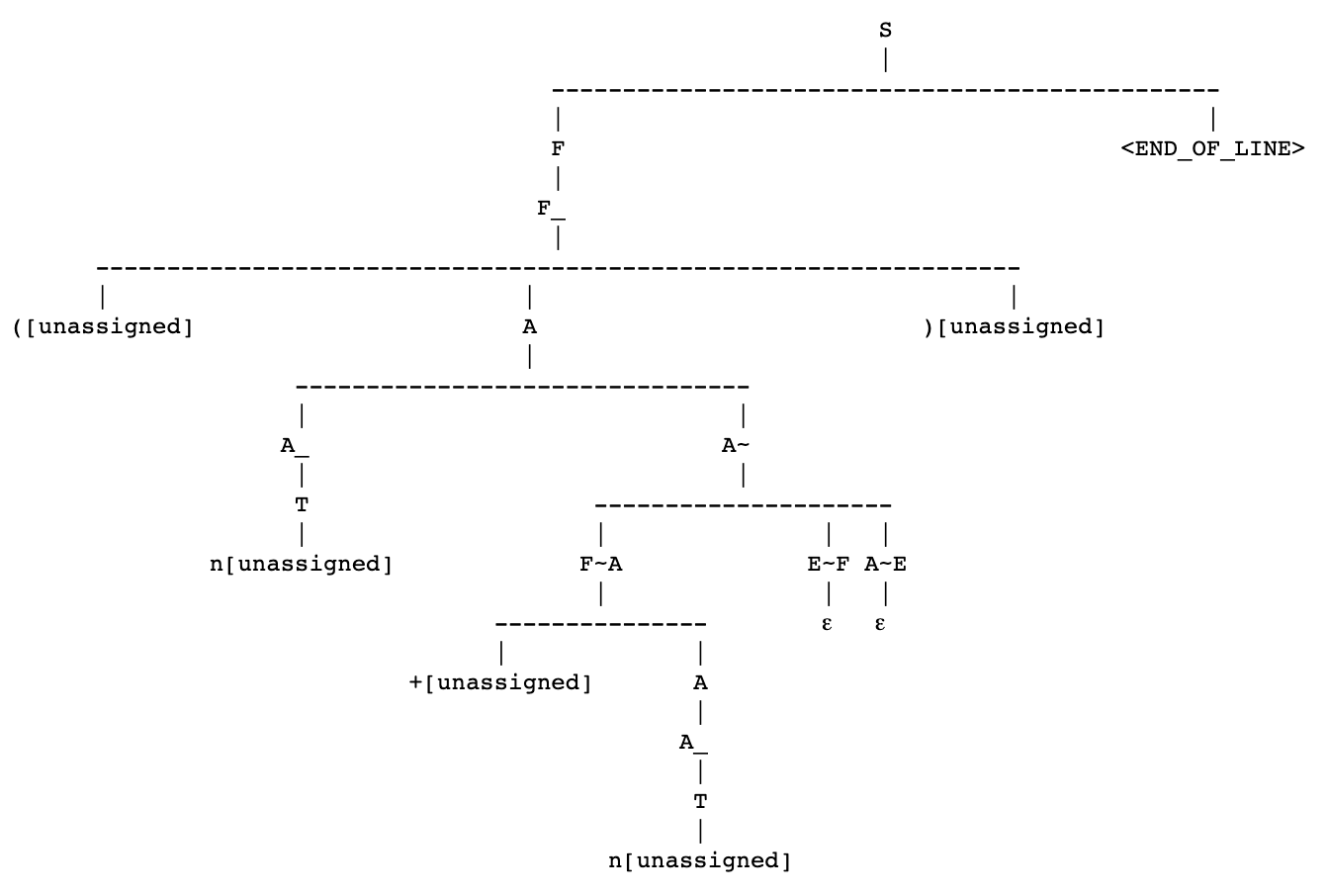} 
\caption{Complex indirect left recursion parse tree}
\label{complex-indirect-left-recursion-parse-tree}
\end{figure}

\noindent Finally we leave the reader with the parse tree that results, shown in figure~\ref{complex-indirect-left-recursion-parse-tree}. Note the presence of the nodes corresponding to the vacuous definitions in some of the indirectly repeated rules. We leave the details of how this parse tree can be rewritten for the next section.

%% file: rewriting.tex
\section{Rewriting the parse trees}

The parse tree shown in figure~\ref{complex-indirect-left-recursion-parse-tree} at the end of the last section is far from ideal in practice. Converting it to the parse tree that would have resulted had there been no need to rewrite the rules can be recovered by a series of relatively simple rewrites, however. We describe each of these rewrites in turn and show the intermediate parse trees that result at each stage in what follows.

\begin{figure}[H]
\centering
\includegraphics[scale=0.5]{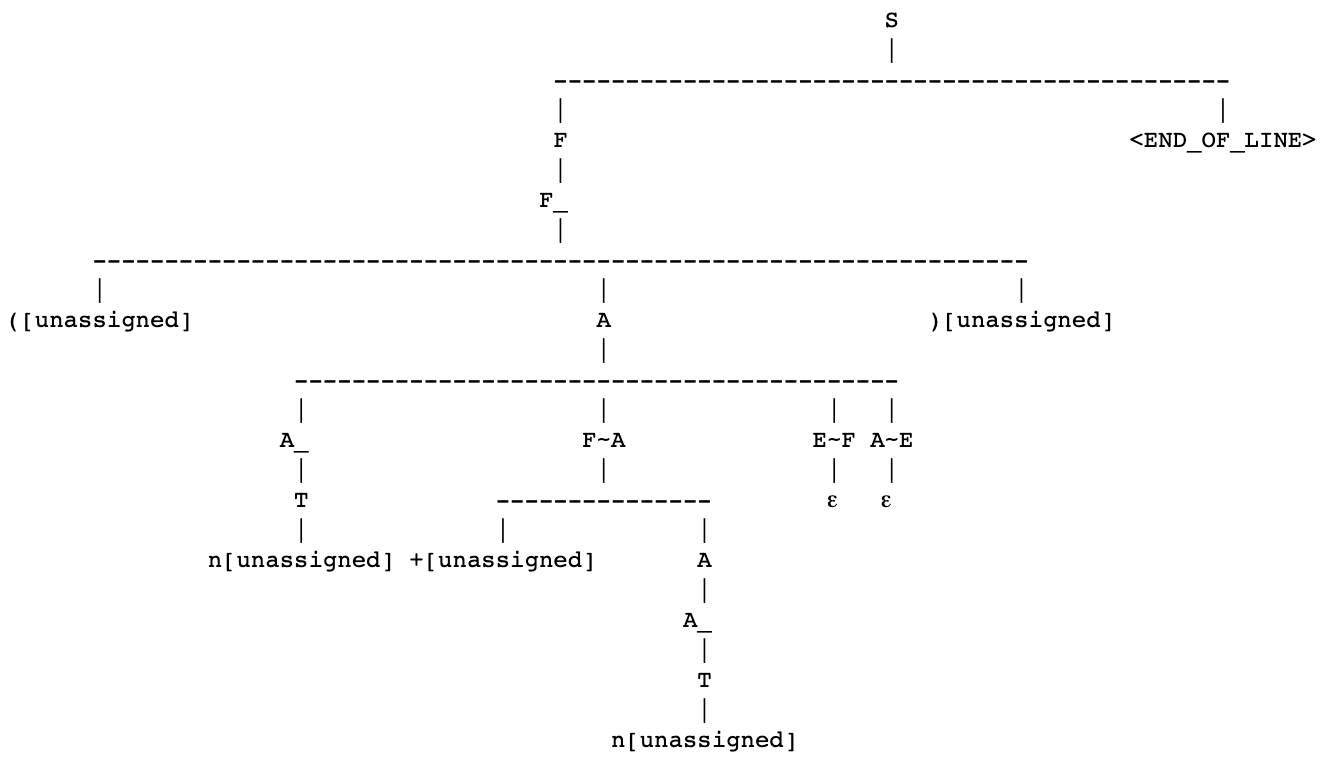} 
\caption{Complex indirect left recursion parse tree with rewritten directly repeated nodes}
\label{complex-indirect-left-recursion-rewritten-directly-repeated-nodes}
\end{figure}

Firstly, the nodes corresponding to the directly repeated rules must be removed. This is easily done but there is a caveat, namely that the order of these nodes must be reversed. This is due to right to left nature of the way the directly repeated rules are executed. Unfortunately, in order to save space, in this case the {\small\texttt{A\textasciitilde}} directly repeated rule is only executed once and therefore we leave this order reversal to the reader's imagination. Moving swiftly on, the resultant parse tree is shown in figure~\ref{complex-indirect-left-recursion-rewritten-directly-repeated-nodes}.

\begin{figure}[H]
\centering
\includegraphics[scale=0.5]{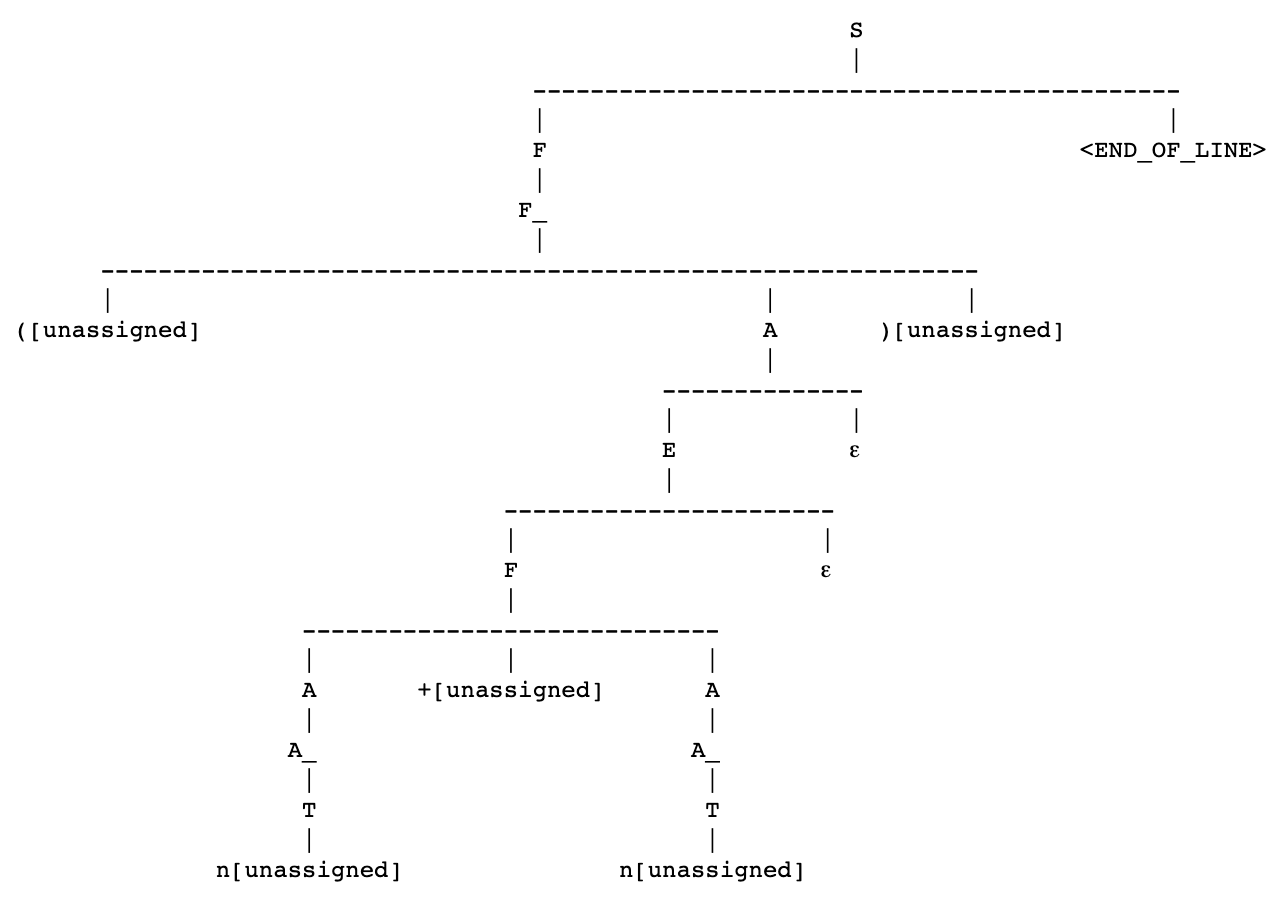} 
\caption{Complex indirect left recursion parse tree with rewritten indirectly repeated nodes}
\label{complex-indirect-left-recursion-rewritten-indirectly-repeated-nodes}
\end{figure}

\noindent Next, the indirectly repeated nodes are rewritten. This is a slightly more convoluted process. Not only is the indirectly repeated node renamed but another node is created and the child nodes are shuffled around. Perhaps the best thing is to simple point the reader to the result in figure~\ref{complex-indirect-left-recursion-rewritten-indirectly-repeated-nodes}. The need to retain the vacuous definitions that result in the epsilon nodes should now be clear. Without them their parent indirectly repeated nodes would not be present in the rewritten parse tree and there would therefore be no opportunity to carry out the necessary rewrites, which create the nodes corresponding to the rewritten rules.\footnote{The author is well aware that the title of the paper should preclude any use of vacuous definitions and epsilon nodes in the algorithm. However, the graph theoretic approach that was finally adopted required them and therefore we lived with the irony rather than inventing another name for a similar usage.}

\begin{figure}[H]
\centering
\includegraphics[scale=0.5]{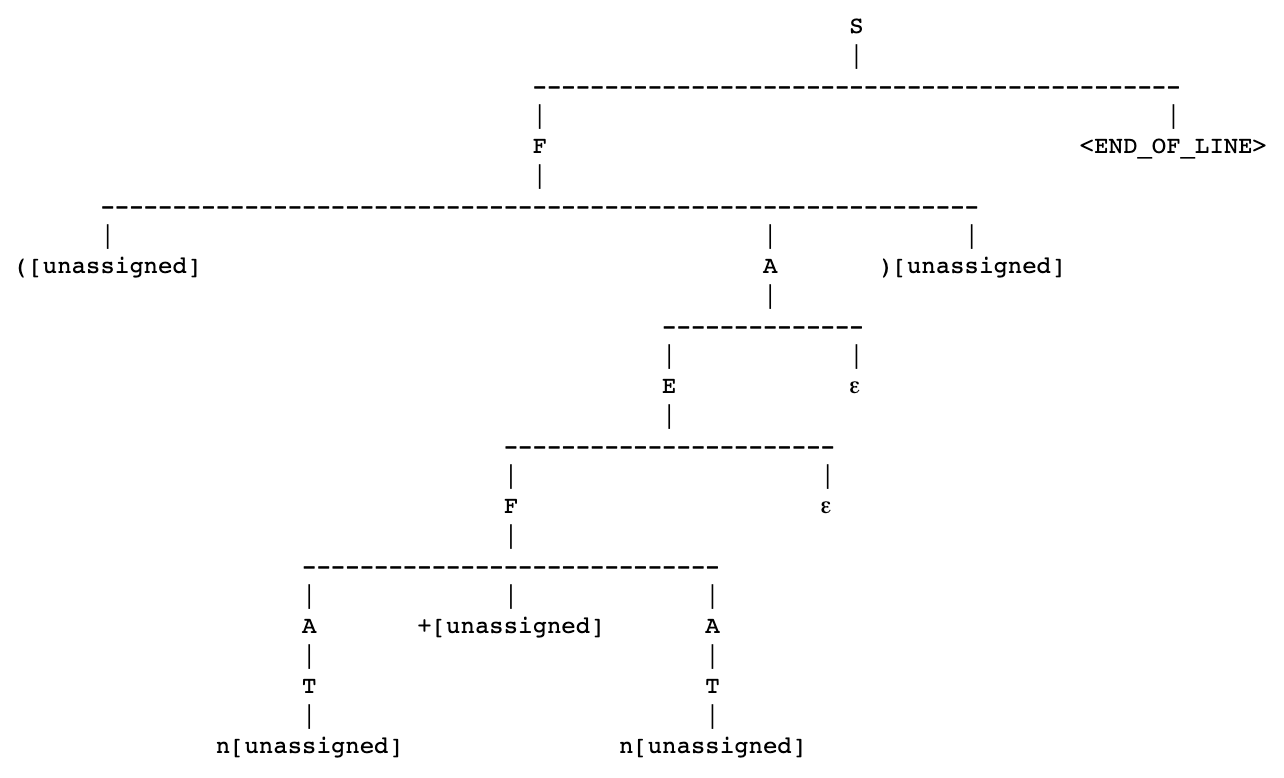} 
\caption{Complex indirect left recursion parse tree with rewritten reduced nodes}
\label{complex-indirect-left-recursion-rewritten-reduced-nodes}
\end{figure}

\begin{figure}[H]
\centering
\includegraphics[scale=0.5]{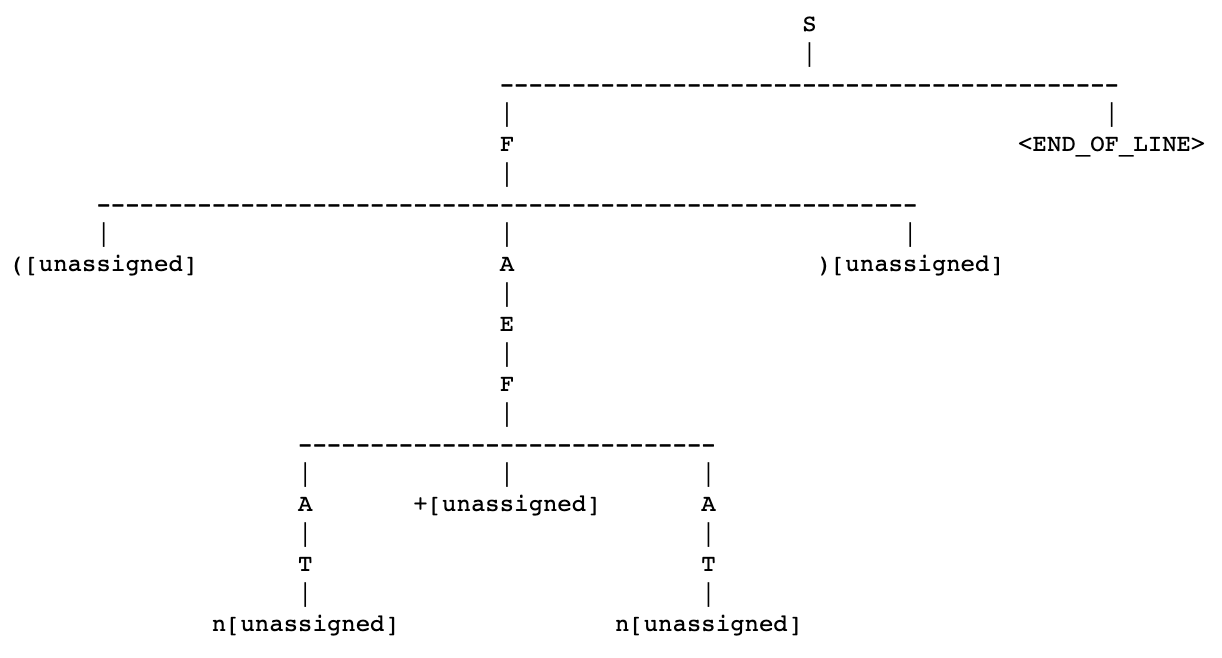} 
\caption{Complex indirect left recursion parse tree with removed epsilon nodes}
\label{complex-indirect-left-recursion-removed-epsilon-nodes}
\end{figure}

\noindent The next step is to rewrite the reduced nodes. If the reduced node has the same underlying name as its parent then it is simply removed. If, on the other hand, the two names differ, then it is simply renamed to the base name.  The resulting parse tree shown in figure~\ref{complex-indirect-left-recursion-rewritten-reduced-nodes}.  Finally the epsilon nodes themselves are removed. The pleasing and final result is shown in figure~\ref{complex-indirect-left-recursion-removed-epsilon-nodes}.

%% file: conclusions.tex
\section{Conclusions}

We have devised a new algorithm for eliminating left recursion in both its direct and indirect forms. We claim that this algorithm is superior to the standard algorithm because it is selective rather than preventative, that is, it eliminates left recursion where it finds it, otherwise leaving the rules intact. The one caveat with this approach is that the algorithm cannot handle certain kinds of first parts in left recursive definitions. However, we assure the reader that this is a minor encumbrance and does not significantly affect the utility of the algorithm in practice.  A JavaScript implementation can found found here~\cite{Occam:Grammar-Utilities}.